\documentclass[aps,prb,reprint,superscriptaddress,floatfix,titlepage]{revtex4-1}
\usepackage{graphicx}
\usepackage[english]{babel}

\begin{document}
\title{Blue-Light-Emitting Color Centers in High-Quality Hexagonal Boron Nitride}
\author{Brian Shevitski}
\affiliation{Department of Physics, University of California at Berkeley, Berkeley, CA 94720, U.S.A.}
\affiliation{Materials Sciences Division, Lawrence Berkeley National Laboratory, Berkeley, CA 94720, U.S.A.}
\affiliation{Kavli NanoEnergy Sciences Institute at the University of California at Berkeley and the Lawrence Berkeley National Laboratory, Berkeley, CA 94729, U.S.A.}
\affiliation{The Molecular Foundry, Lawrence Berkeley National Laboratory, Berkeley, CA 94720, U.S.A.}
\author{S. Matt Gilbert}
\affiliation{Department of Physics, University of California at Berkeley, Berkeley, CA 94720, U.S.A.}
\affiliation{Materials Sciences Division, Lawrence Berkeley National Laboratory, Berkeley, CA 94720, U.S.A.}
\affiliation{Kavli NanoEnergy Sciences Institute at the University of California at Berkeley and the Lawrence Berkeley National Laboratory, Berkeley, CA 94729, U.S.A.}
\author{Christopher T. Chen}
\affiliation{The Molecular Foundry, Lawrence Berkeley National Laboratory, Berkeley, CA 94720, U.S.A.}
\author{Christoph Kastl}
\affiliation{The Molecular Foundry, Lawrence Berkeley National Laboratory, Berkeley, CA 94720, U.S.A.}
\affiliation{Walter-Schottky-Institute and Physik Department, Technical University of Munich, Garching, 85748, Germany.}
\author{Edward S. Barnard}
\author{Ed Wong}
\author{D. Frank Ogletree}
\affiliation{The Molecular Foundry, Lawrence Berkeley National Laboratory, Berkeley, CA 94720, U.S.A.}
\author{Kenji Watanabe}
\author{Takashi Taniguchi}
\affiliation{Advanced Materials Laboratory, National Institute for Materials Science, 1-1 Namiki, Tsukuba,305-0044, Japan}
\author{Alex Zettl}
\email{\tt azettl@berkeley.edu}
\affiliation{Department of Physics, University of California at Berkeley, Berkeley, CA 94720, U.S.A.}
\affiliation{Materials Sciences Division, Lawrence Berkeley National Laboratory, Berkeley, CA 94720, U.S.A.}
\affiliation{Kavli NanoEnergy Sciences Institute at the University of California at Berkeley and the Lawrence Berkeley National Laboratory, Berkeley, CA 94729, U.S.A.}
\author{Shaul Aloni}
\email{\tt saloni@lbl.gov}
\affiliation{The Molecular Foundry, Lawrence Berkeley National Laboratory, Berkeley, CA 94720, U.S.A.}
\date{\today}
\begin{abstract}
Light emitters in wide band gap semiconductors are of great fundamental interest and have potential as optically addressable qubits. Here we describe the discovery of a new color center in high-quality hexagonal boron nitride (h-BN) with a sharp emission line at 435 nm. The emitters are activated and deactivated by electron beam irradiation and have spectral and temporal characteristics consistent with atomic color centers weakly coupled to lattice vibrations. The emitters are conspicuously absent from commercially available h-BN and are only present in ultra-high-quality h-BN grown using a high-pressure, high-temperature Ba-B-N flux/solvent, suggesting that these emitters originate from impurities or related defects specific to this unique synthetic route. Our results imply that the light emission is activated and deactivated by electron beam manipulation of the charge state of an impurity-defect complex.
\end{abstract}

\maketitle

\section{Introduction}
Luminescent defects in wide band gap semiconductors are of great importance for both fundamental physics and future technological applications. Many of these defects are single photon emitters (SPEs), a likely component of next-generation information technologies, especially quantum cryptography\cite{lo_secure_2014,scarani_security_2009} and information processing\cite{kok_linear_2007,obrien_photonic_2009,northup_quantum_2014}. SPEs embedded in solid state systems are particularly significant for widespread adoption of these emerging technologies as they offer a promising route toward scalable deployment of new integrated quantum circuits. The diamond nitrogen-vacancy (NV) center has been the leading candidate for solid-state SPE applications because it can easily be manipulated and readout at room temperature using existing optical methods\cite{kurtsiefer_stable_2000,neumann_multipartite_2008}. Due to the technical difficulty of synthesizing and fabricating diamond-based devices, greater attention has been placed on finding new solid-state SPE systems\cite{castelletto_silicon_2014,morfa_single-photon_2012} with particular emphasis placed on 2D material systems, especially $sp^2$-bonded hexagonal boron nitride (h-BN) )\cite{martinez_efficient_2016,tran_quantum_2016,tran_robust_2016,tran_quantum_2016-1,bourrellier_bright_2016}.
\par
h-BN or ``white graphite" has been of great interest to the nanoscience community over the last several decades, in part because it is isostructural to graphite and forms many of the same types of nanostructures as $sp^2$-bonded carbon, but with different electronic and thermodynamic properties\cite{chopra_boron_1995,rubio_theory_1994,novoselov_two-dimensional_2005,nagashima_electronic_1995-1,golberg_octahedral_1998}. h-BN is especially important to the expanding study of 2D materials because it is atomically flat, inert, and electrically insulating, making it an ideal substrate for testing new physics in low-dimensional materials\cite{dean_boron_2010}. Recently it has been shown that the local structure of h-BN can be controlled via electron irradiation\cite{pham_formation_2016,gilbert_fabrication_2017} and synthetic methods\cite{gilbert_alternative_2019}, allowing for additional material control.
\par
In this communication we report the discovery of a new color center in high- quality h-BN, activated and characterized by an electron beam. We show that these emitters are highly localized, are spectrally pure, and have an emission signature indicative of weak lattice coupling at room temperature. These properties make the new color center a promising candidate for future applications in quantum information science.
\par
\section{Experimental}
Ultra-high-quality h-BN crystals used in this study (generally accepted as the best substrates available for sensitive device applications and scanning probe measurements) are synthesized using a Ba-B-N solvent precursor at high temperature and high pressure at the National Institute for Materials Science (NIMS) by Watanabe and Taniguchi\cite{taniguchi_synthesis_2007}. We refer to this material throughout the text as NIMS-BN. Millimeter size crystallites are mechanically exfoliated using blue wafer dicing tape and transferred onto p$++$ silicon substrates (0.001-100 $\Omega$-cm) with a few nm thick native oxide layer. Samples of commercially available h-BN (Alfa Aesar 040608) are similarly exfoliated and transferred, and then annealed at 850 $^{\circ}$C in argon at one Torr for one hour before characterization.
\begin{figure*}[ht]
\includegraphics[width=0.8\textwidth]{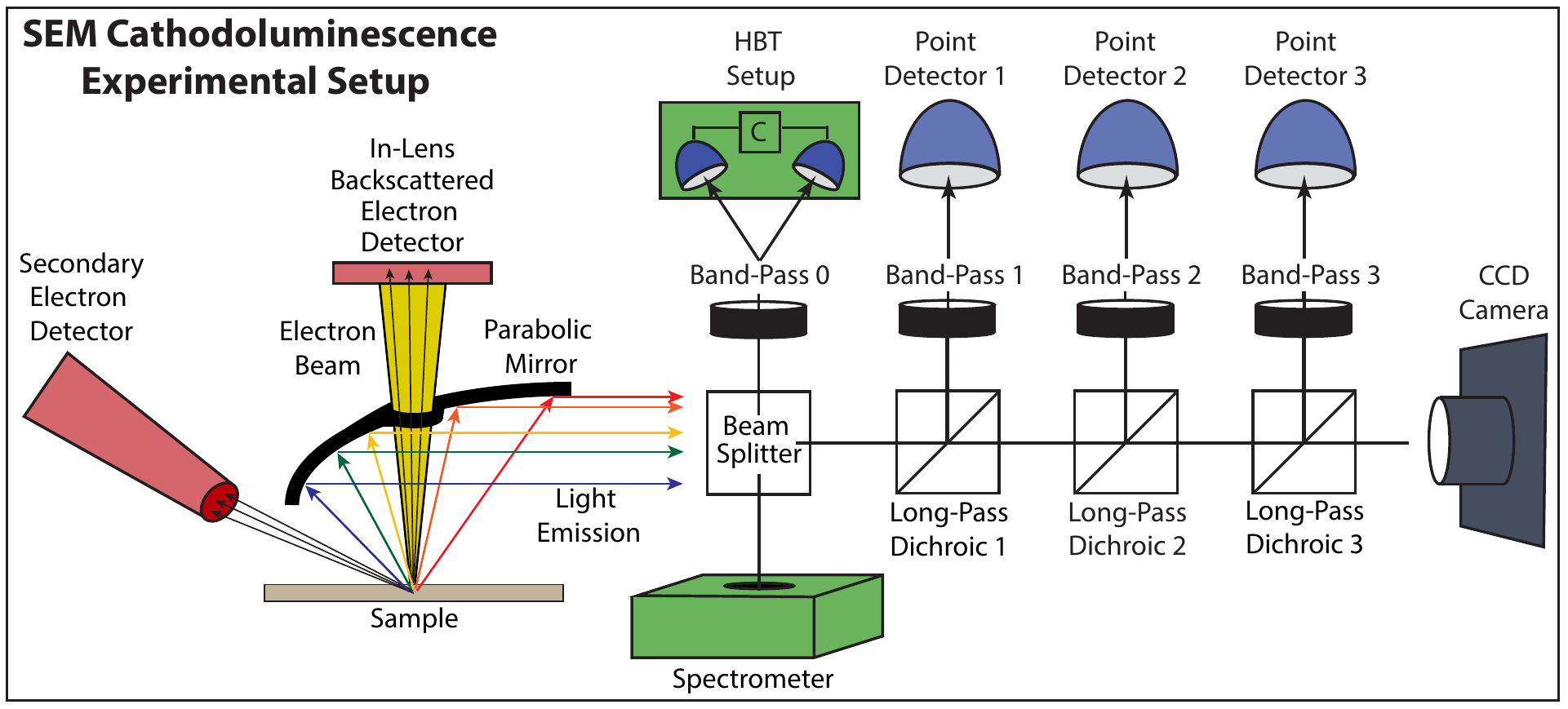}
\caption{\label{fig:CLSetup} Experimental setup for measuring cathodoluminescence (CL) in the scanning electron microscope (SEM). The electron beam excites the sample and causes it to fluoresce. The resulting light is then used for hyperspectral imaging by a spectrometer (slow acquisition, low signal to noise, high spectral resolution) and point detectors (fast acquisition, high signal to noise, low spectral resolution). A Hanbury-Brown Twiss (HBT) apparatus is used to measure the second order coherence function ($g^2\left(\tau\right)$) of the light emission.}
\end{figure*}
\par
We use CL in a scanning electron microscope (SEM) to activate and characterize light emission from the h-BN. CL measurements are performed using a home-built SEM CL system shown in FIG. \ref{fig:CLSetup} The system is built around a Zeiss Gemini Supra 55 VP-SEM operated at accelerating voltages between 2-10 keV with beam currents in the 100-1500 pA range. Light emission from the sample is collected by a parabolic mirror and directed down an optical path for characterization. Synchronous data from SEM (secondary electron and in-lens detectors) and optical (photon counting point detectors and spectrometer) channels are acquired using the Molecular Foundry ScopeFoundry software\cite{durham_scanning_2018}. All experiments are conducted at room temperature. 
\par
As the electron beam is rastered across the sample, the light emission from each point of the scan can be coupled into an optical fiber (Thorlabs FG200UEA) using a UV-enhanced parabolic aluminum reflector and recorded using a spectrometer (Princeton Instruments SP2300i) with CCD camera (Andor 970-UVB) to capture the spectral distribution of the light at each pixel, resulting in a three-dimensional data set we refer to as a spectral image (SI). The spectra are not intensity corrected for the wavelength dependent efficiency of the spectrometer grating and CCD camera.
\par
Alternatively, the light can be directed through a series of dichroic mirrors and bandpass filters to an array of photon counting photomultiplier tube point-detectors (Hamamatsu H7360-01, Hamamatsu H7421-40, and Hamamatsu H7421-50), resulting in intensity images of well-defined wavelength bands. We refer to such data throughout the text as bandpass (BP) images.
\par
Time-correlation of the emitted light can be measured by coupling to a Hanbury-Brown-Twiss (HBT) setup. The arrival times of photons at the detectors in both arms of the apparatus are recorded with 50 ps resolution and a coincidence histogram as a function of delay time between the two detectors is made. The raw coincidence histogram is then normalized by the number of coincidences at long delay times. A background correction is performed using the signal to background ratio estimated independently for each measurement\cite{brouri_photon_2000}, resulting in a measurement of the second-order auto-correlation function ($g^2\left(\tau\right)$) of the emitted light (See Supplemental for details).
\par
All data analysis is performed using common open-source packages in the Python programming language. Multivariate statistical analysis (MVA) of hyperspectral images is carried out using the HyperSpy Python package\cite{francisco_de_la_pena_hyperspy/hyperspy:_2018}.
\section{Results}
\begin{figure*}[ht]
\centering
\includegraphics[width=0.97\textwidth]{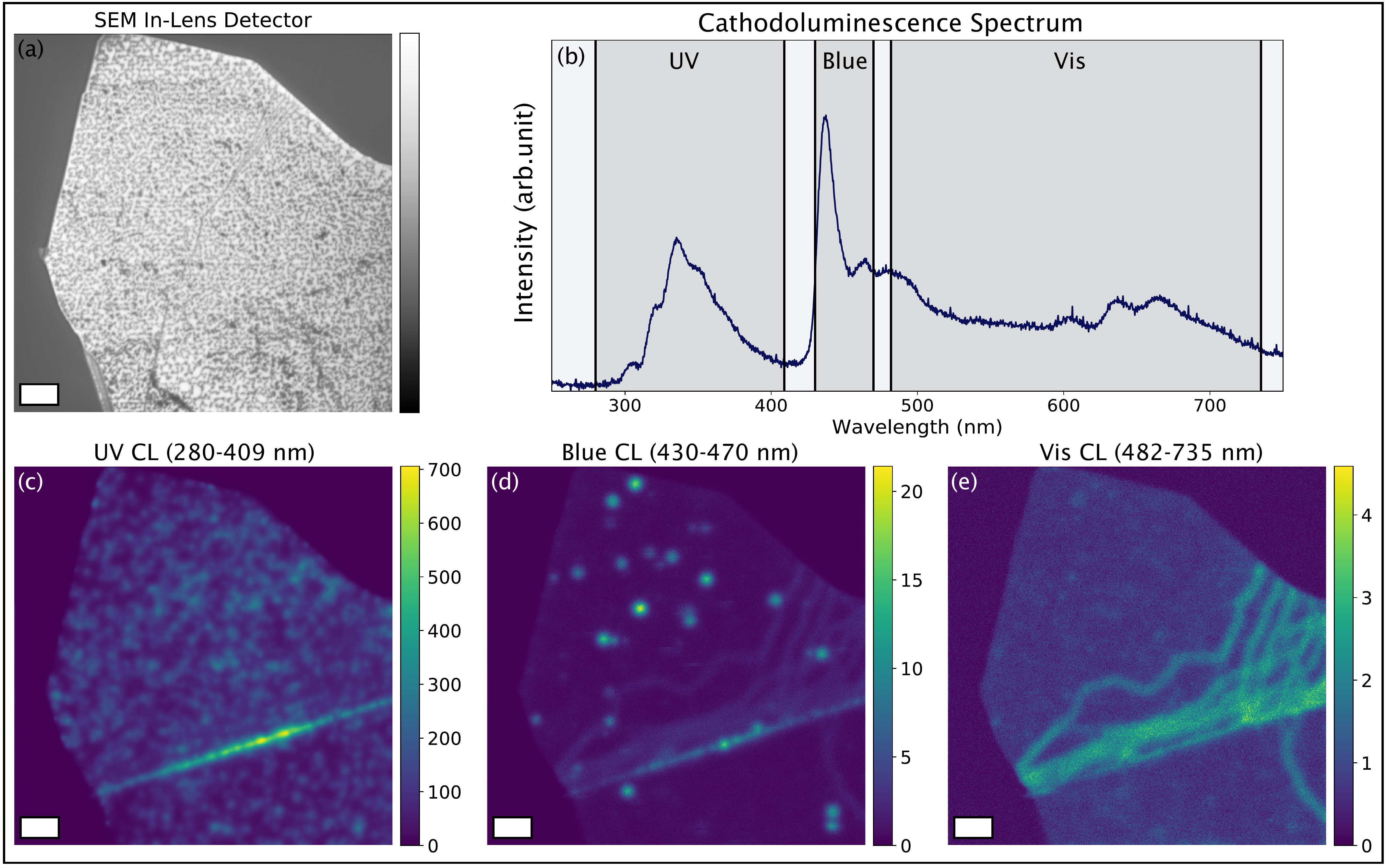}
\caption{\label{fig:CLOverview} Overview of light emission from h-BN. An SEM image of a flake of h-BN is shown in (a), scale bar is 1 $\mu$m. Panel (b) shows the mean Cathodoluminescence (CL) response from the sample, with the three main spectral bands of interest indicated by vertical lines and shaded regions. Panels (c-e) show the light emission from the flake in each of these 3 spectral bands (wavelength range listed above each image, intensity scale in counts/pixel). Panel (c) shows UV light emission from a high density of point-like emitters as well as bright emission from a large line-like feature. The intensity in the Blue band arises mostly from the new point-like emitters, shown in Panel (d). Panel (e) shows the spatial distribution of intensity from the green to red region of the spectrum (Vis). Emission in this band arises mostly from extended defects in the crystal. The images are generated by averaging 17 consecutive scans, each with a pixel dwell time of 28.6 $\mu s / \text{pixel}$.}
\end{figure*}
\par
We first describe results from NIMS-BN. In order to activate and characterize individual emitters, we initially identify large (lateral size of 10-100 $\mu$m) flakes of h-BN using the raster scan images from the SEM electron detectors, shown in FIG. \ref{fig:CLOverview}(a). Optical and atomic force microscopy (AFM) observations of our samples indicate that the flakes generally have thicknesses of tens to hundreds of nanometers. Relevant spectral bands for BP imaging are determined by inspection of the average CL spectrum, shown in FIG. \ref{fig:CLOverview}(b), obtained by integrating spectra while continuously scanning a single flake. We isolate the primary spectral regions of interest (indicated by the vertical lines and shaded regions in FIG. \ref{fig:CLOverview}(b)): 280-409 nm (UV), 430-470 nm (Blue), and 485-735 nm (Vis) for BP imaging with appropriate dichroic and bandpass optics.
\par
There are several striking features in each channel of the BP images. FIG. \ref{fig:CLOverview}(c) shows the spatial distribution of the previously reported UV emitters in the h-BN flake\cite{bourrellier_bright_2016}. The emitters have a point-like character and are densely and uniformly distributed across the entire h- BN crystal, with enhanced emission along line-like features that are likely associated with extended line defects\cite{jaffrennou_origin_2007} or strain caused by a wrinkle or fold in the crystallite\cite{proscia_near-deterministic_2018}. FIG. \ref{fig:CLOverview}(e) shows very weak extended features in the Vis band (485-735 nm) as previously reported\cite{jaffrennou_origin_2007}. These features are localized at grain-boundaries and dislocations within the h-BN crystal. No localized, point-like sources of light emission are observed within this band using CL BP imaging.
\par
The bright, point-like features in the Blue BP image between 430-470 nm in FIG. \ref{fig:CLOverview}(d) show the new color centers in NIMS-BN, which are the focus of this study. They appear slightly larger and less dense compared to the UV emitters in the previous panel. Also present is the line-like feature from the UV band, as well as a very weak signal from extended features in the Vis band. The broad spectral character of these extended features results in residual intensity “leaking” into the Blue band.
\begin{figure}[ht]
\includegraphics[width=0.48\textwidth]{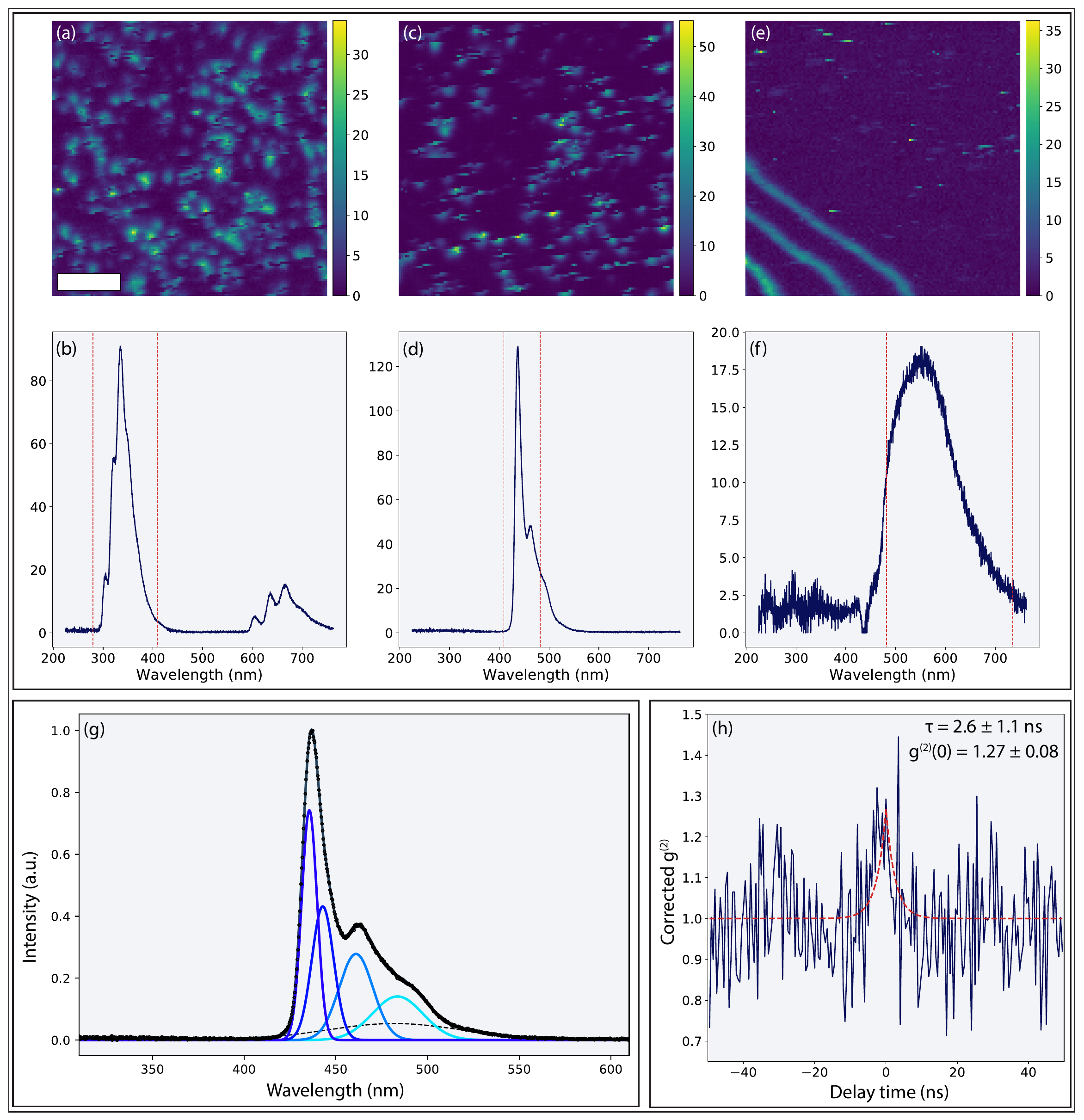}
\caption{\label{fig:HSOverview} Hyperspectral characterization of a dense array of emitters across three spectral regions in h-BN. Hyperspectral CL data are decomposed using Non-Negative Matrix Factorization (NMF) to display the main features of the data. Each image component (top row, scale bar 1 um) shows the spatial distribution of the associated spectral component below (middle row). The red lines on the spectral components in panels (b), (d), and (f) show the edges of the bandpass filters used in bandpass imaging from FIG. \ref{fig:CLOverview}. The spatial (a) and spectral (b) distributions of the first component show point-like UV emitters from 300-400 nm. The peaks between 600-700 nm are artefacts due to the second-order reflections of the UV light from the spectrometer grating. Panels (c) and (d) show the spectral signature of the newly discovered blue color centers between 400-500 nm. Panels (e) and (f) show extended emission features between 500-700 nm, likely caused by extended defects and/or strain. Panel (g) shows a rescaled view of the blue emission component as well as the results of multi-gaussian fitting of the peak. The fitted components show a Zero-Phonon Line (ZPL) followed by several phonon replicas with decreasing amplitude, an indicator of a single-photon emitter weakly coupled to the lattice. The photon-correlation curve in (h) shows a bunching peak, a signature of quantum emission in CL. Fitting to a single exponential decay model gives a lifetime of several nanoseconds, similar to lifetimes observed from SPE in other defects in h-BN.}
\end{figure}
\par
We further characterize electron-stimulated light emission from NIMS-BN using hyperspectral CL imaging by collecting a CL spectrum from each pixel in a raster scan. To isolate the unique spectral signature of each type of emitter we perform a non-negative matrix factorization (NMF) decomposition\cite{arngren_unmixing_2011} of the spectra in the SI into four components that visualize the main features present in the dataset. FIG. 3(a-f) show the results of the decomposition. Each image (also called the decomposition loading or decomposition weight) shows the relative abundance of each associated spectral component (also called the decomposition factor) below. The vertical red lines in FIG. \ref{fig:HSOverview}(a), \ref{fig:HSOverview}(c), and \ref{fig:HSOverview}(e) show the pass band used for BP imaging in FIG. \ref{fig:CLOverview}(c), \ref{fig:CLOverview}(d), and \ref{fig:CLOverview}(e), respectively. 
\par
The first component (see Supplemental) is a spatially uniform background that reflects dark counts, noise, and non-localized light emission. The component shown in FIG. \ref{fig:HSOverview}(e) and \ref{fig:HSOverview}(f) shows the light emission of extended line defects, as well as a small number of highly localized features with appreciable intensity that are consistent with previous photoluminescence (PL) studies\cite{martinez_efficient_2016,tran_quantum_2016} of SPE in h-BN. FIG. \ref{fig:HSOverview}(a) shows a dense collection of point-like emitters, similar to the UV BP image in Fig. \ref{fig:CLOverview}(b). The spectral features in the component in FIG. \ref{fig:HSOverview}(b)) are a close match to UV SPEs in h-BN seen in previous CL studies\cite{bourrellier_bright_2016}. 
\par
The most striking feature, shown in FIG. \ref{fig:HSOverview}(c) and \ref{fig:HSOverview}(d), closely resembles the spectral and spatial signatures of a typical color center in a wide bandgap semiconductor\cite{martinez_efficient_2016,tran_robust_2016}. Specifically, they are highly localized, spectrally sharp, and the first peak is followed by additional spectral features shifted by tens of meV that are interpreted as evidence of electron-phonon coupling. This component is investigated further in FIG. \ref{fig:HSOverview}(g) using multi-Gaussian fitting of the spectral component. Fitting reveals a series of peaks, decreasing in intensity and increasing in width with increasing wavelength. The spectrum is dominated by a sharp, well-defined zero phonon line (ZPL), centered at 436 nm (2.84 eV), with a FWHM of 10 nm (65 meV), contributing a spectral weight of $\sim 26 \%$, followed by several phonon replicas at 443, 461, and 484 nm (2.80, 2.69, and 2.56 eV). There is also a small, broad, background component centered at 482 nm (2.57 eV). In contrast to the UV emission described above (where the phonon replicas have higher intensity than the ZPL), the relatively high intensity of the ZPL compared to the phonon replicas suggest that this emitter’s coupling to the lattice is significantly weaker, a desirable quality for possible applications in future quantum information technologies. The density of emitters in FIG. \ref{fig:HSOverview}(c) is higher than in the BP image in FIG. \ref{fig:CLOverview}(d) and is related to the higher electron dose required for hyperspectral imaging, which will be discussed later.
\par
Spectral information from the hyperspectral imaging can be used to perform time-correlation measurements of light from the blue emitters. A bandpass filter is selected that covers a large portion of the emission peak from FIG. \ref{fig:HSOverview}(d) in order to measure the second-order autocorrelation function using HBT intensity interferometry. FIG. \ref{fig:HSOverview}(h) shows the result of this measurement from an ensemble of emitters (See Supplemental for experimental details). 
\par
\begin{figure*}[ht]
\includegraphics[width=0.98\textwidth]{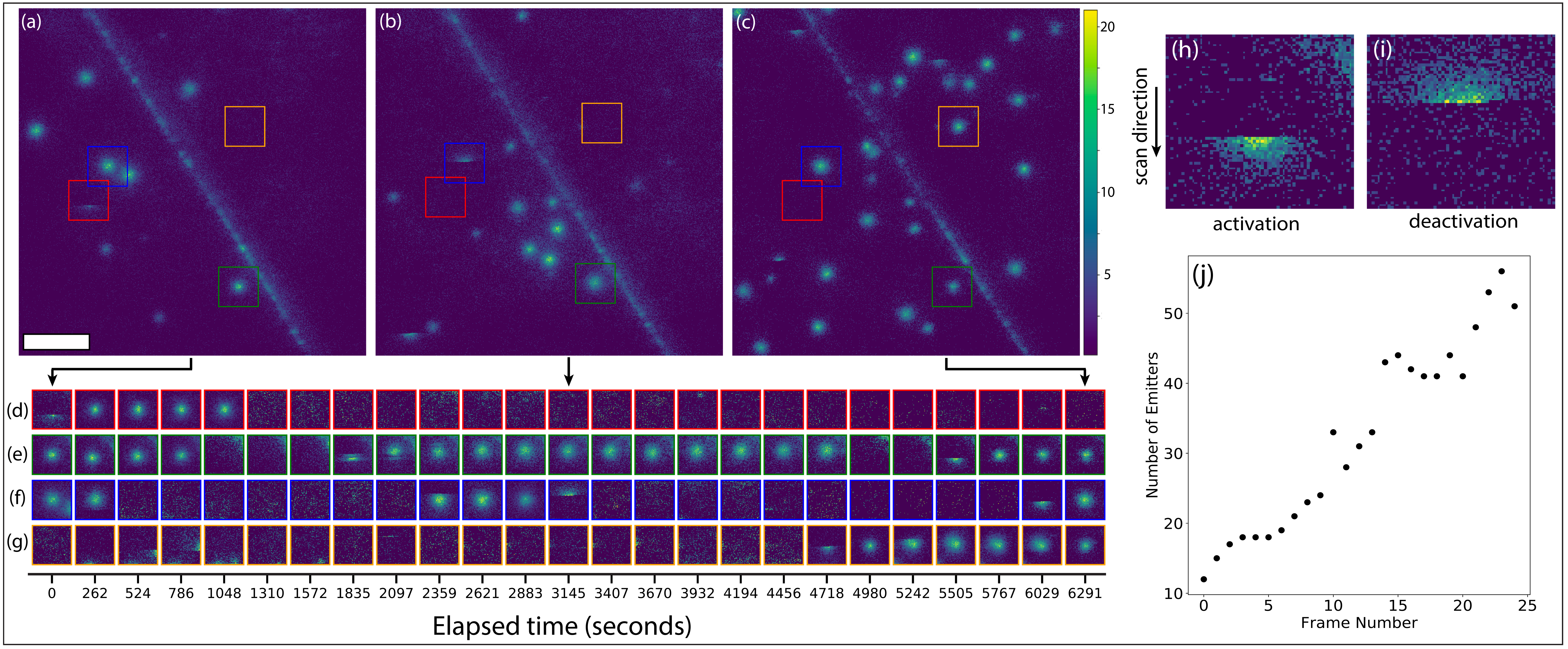}
\caption{\label{fig:Dynamics} Overview of switching behavior in hBN emitters. A series of CL bandpass images (409- 482 nm) are taken over ~1.5 hours at ~4.5 minutes per image, with each time point corresponding to a single image scan with an electron beam dose of $2.3 \times 10^{7} \; e^{-}/\text{\normalfont\AA}^2$. The top row shows the first, middle, and final images in the series at full spatial resolution (scale bar is 500 nm, color scale in units of counts/pixel). The bottom row shows the dynamic behavior of 4 separate 300 nm regions (indicated by the colored boxes in the top panel) over the entire time series. The emitters exhibit clear switching behavior, turning on and off between frames of the time series. Examples of emitters that suddenly turn on or off as the beam is passing over are shown in panels (h) and (i), respectively. Panel (j) shows that the number of emitters in each frame grows roughly linearly with electron beam dose.}
\end{figure*}
The majority of previous studies of light emission in h-BN have measured $g^{(2)}_{PL}\left(\tau\right)$ using PL. In stark contrast, $g^{(2)}_{CL}\left(\tau\right)$, measured in this study, exhibits a bunching peak as opposed to an anti-bunching dip. This behavior, which has been explored previously\cite{feldman_colossal_2018,meuret_photon_2017}, is attributed to simultaneous excitation of multiple color centers by the electron beam. In PL studies of defects in wide band gap semiconductors, such as SPEs in h-BN and NV centers in diamond, the excitation energy is typically less than the band-gap of the material, resulting in the production of a single e-h pair per photon. In CL, the excitation energy is much higher than the band-gap of the material, resulting in the excitation of many e-h pairs. A semi-empirical relation predicts that approximately $N_{e-h} = \frac{E_0}{3E_g}$ e-h pairs are excited per incident electron\cite{klein_bandgap_1968}, where $E_0$ is the beam energy and $E_g$ is the band gap of h-BN. For $E_g = 6 \; \text{eV}$ and $E_0 = 3 \; \text{keV}$, $N_{e-h} = 111$.
\par
Previous work\cite{meuret_photon_2015} has shown that if the bunching peak of the $g^{(2)}_{CL}\left(\tau\right)$ function can be attributed to simultaneous excitation of an ensemble of emitters, that the lifetime of the defect state can be extracted similar to an analogous PL measurement of the lifetime from the anti-bunching dip. An exponential fit to the time correlation data using $g^{(2)}\left(\tau\right) = 1 + \text{a} \exp\left(-\left|\tau\right|/\tau_{lifetime}\right)$, with a and $\tau_{lifetime}$ as free parameters, gives a lifetime of $\tau_{lifetime} = 2.6 \; \text{ns}$ and $g^{(2)}\left(0\right) = 1.27$. This lifetime is close in value to previous measurements of SPEs in h-BN\cite{bourrellier_bright_2016,meuret_photon_2015,martinez_efficient_2016,tran_robust_2016,tran_quantum_2016,tran_quantum_2016-1}. The $g^{(2)}\left(0\right)$ value is quite low, but it has been shown that at high current, the bunching effect becomes washed out, resulting in a decreased apparent value of $g^{(2)}\left(0\right)$. While the observed bunching peak and extracted fit parameters are not incontrovertible proof that the new color center is a quantum emitter, the presence of this peak is consistent with the newly discovered blue emitter being a potential single photon source.
\par
There are two striking differences between the Blue emitters shown in FIG. \ref{fig:CLOverview}(d) and those shown in FIG. \ref{fig:HSOverview}(c) that are associated with the higher electron dose required to acquire an SI compared to a BP image. First, the number and density of features is higher in the SI, indicating that we are creating new emitters by electron irradiation. Second, a variation in the shape of individual emitters appears. The emitters are round and symmetric in the BP image in FIG. \ref{fig:CLOverview}(d), while many emitters appear to have a truncated shape in the SI component in FIG. \ref{fig:HSOverview}(c). This truncation, discussed below, is associated with a sudden activation or deactivation event while the beam is over an emitter. 
\par
To further investigate this, we acquire a 2.5-hour time series of long scan time (262 seconds per image) BP images over a flake of h-BN. Each image corresponds to an exposed dose of $2.3 \times 10^{7} \; e^{-}/\text{\normalfont\AA}^2$ per image, with a total exposed dose of $5.7 \times 10^{8} \; e^{-}/\text{\normalfont\AA}^2$. Time-series data are aligned and registered using template matching and cross-correlation to correct for sample drift during the experiment. Individual emitters were automatically identified in each frame of the time series using the difference of Gaussian blob finding algorithm. 50x50 pixel regions around emitters were extracted and image feature vectors were calculated using the PCA weights of each image. Finally, false positives were removed by inspection of the output of k-means clustering on the feature vectors.   
\par
FIG. \ref{fig:Dynamics} summarizes the results of this experiment. The top row shows the first (FIG. \ref{fig:Dynamics}(a)), middle (FIG. \ref{fig:Dynamics}(b)), and last (FIG. \ref{fig:Dynamics}(c)) images from the time series at full spatial resolution. The colored boxes in the full resolution images indicate regions where we have cropped the data and displayed the entire time series in a 300 nm region around four emitters (FIG. \ref{fig:Dynamics}(d-g)). FIG. \ref{fig:Dynamics}(d) shows an emitter that is suddenly activated by the electron beam (indicated by the truncated disc shape of the emitter), remains in the emissive state for 5 frames, then suddenly switches off for the remainder of the time series. FIG. \ref{fig:Dynamics}(e-f) show emitters that switch on and off several times throughout the scan. Some of these activation/deactivation events appear as truncated discs (shown in detail in FIG. \ref{fig:Dynamics}(h-i)), while some appear as the sudden appearance or disappearance of a bright spot from one frame to another. FIG. \ref{fig:Dynamics} illustrates that not only are the emitters activated and deactivated by the electron beam, but they also disappear and reappear in identical spatial locations (within the accuracy of our measurement)
\par
We note that we have also attempted characterization of these emitters using PL. NIMS-BN samples on indexed Si substrates are seeded with emitters using the SEM, characterized using CL, and then transferred to a CW PL setup using 1 mW 349 nm, 10 mW 405 nm, and 30 mW 532 nm excitation focused to a diffraction limited spot. No emission is observed in the Uv-Vis range using the available excitation energies and intensities within reasonable integration times. This fact, in conjunction with the activation/deactivation behavior observed under the electron beam, hints that the origin of this emission is associated with the charge state of an electron irradiated point defect, similar to SPE in other semiconductor systems\cite{hauf_chemical_2011,siyushev_optically_2013,grotz_charge_2012}. 
\par
Furthermore, we note that the Blue 435 nm emission is not observed using PL or CL in any samples of commercial h-BN powder. This implies that the new emission may be closely related to the unique synthetic origin of NIMS-BN.
\par
\section{Discussion}
We consider possible mechanisms for the new emission. We immediately rule out the direct creation of defects via knock-on damage and electron-beam-induced heating. The electron energy threshold for knock-on damage in h-BN is in the range of 70-80 keV\cite{kotakoski_electron_2010,zobelli_electron_2007}, far greater than the 1-10 keV energy range of the SEM beam. Electron-beam-induced heating is unlikely due to the high thermal conductivity of h-BN along with the relatively low current of the electron beam. 
\par
The change in temperature of the sample can be estimated\cite{reimer_scanning_2013} assuming that energy from the electron beam is uniformly deposited in a sphere of radius R using $\Delta \text{T}= 3\text{IV}f/2\pi\text{R}\kappa$, where I is the beam current, V is the beam accelerating voltage, $f$ is the fraction of incident energy that is absorbed, and $\kappa$ is the thermal conductivity of h-BN (600 W/m K). Assuming that 100$\%$ of the incoming power is absorbed in a spherical interaction volume of radius R = 30 nm, a 2 keV electron beam with 1 nA of current causes a temperature increase of $\Delta \text{T} = 0.05 \; \text{K}$. This value would be even lower at increased accelerating voltage\cite{donolato_analytical_1981} because the interaction size scales approximately as $\text{V}^{1.75}$. Furthermore, due to the finite thickness of an h-BN flake, the fraction of energy from the beam deposited into the sample decreases at higher beam energies. 
\par
\begin{figure}[ht]
\includegraphics[width=0.48\textwidth]{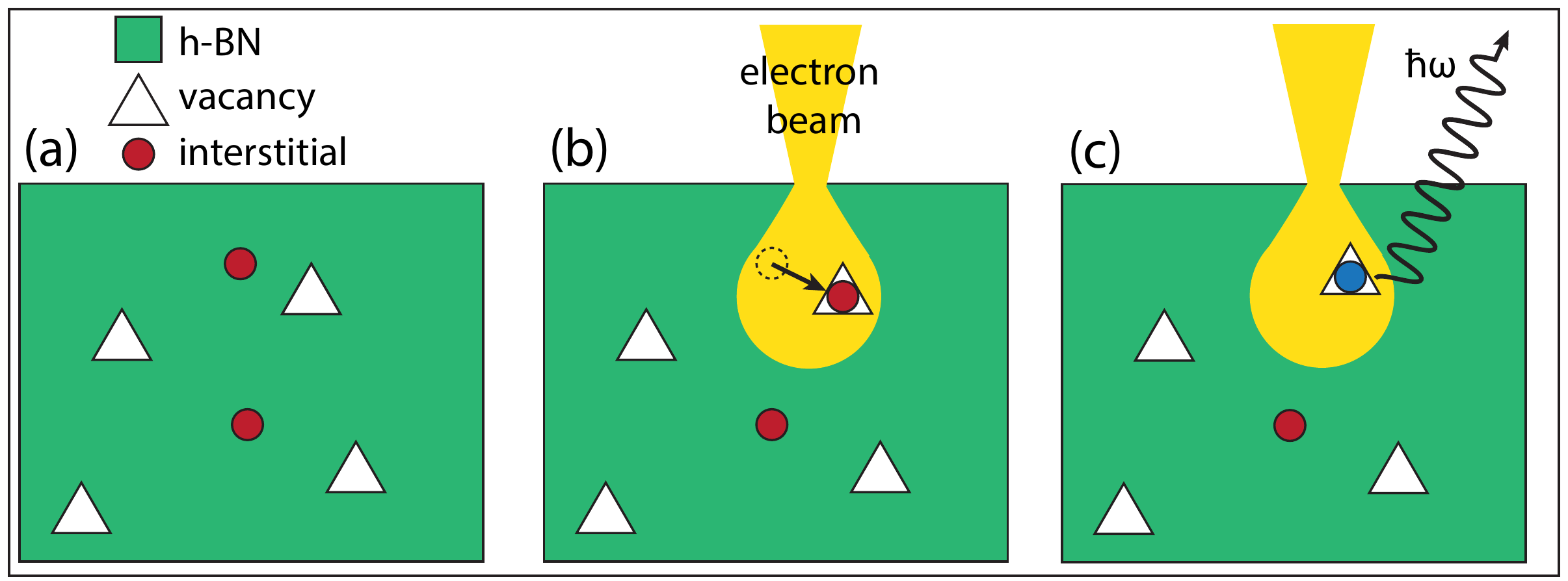}
\caption{\label{fig:Cartoon} Proposed model for formation and activation of blue color centers in NIMS-BN. The as-synthesized material has some intrinsic density of interstitial impurities and lattice vacancies, as shown in (a). When stimulated by the electron beam, the impurity atoms are driven into the vacancies and due to the energy provided by the beam a substitutional defect complex is formed. The charge state of the new defect is changed by the beam causing it to emit photons. The charge state of the defect is sensitive to the intense stimulus from the electron probe and can jump between emissive and non-emissive states, resulting in a blinking effect.}
\end{figure}

We propose that the origin of this new emitter is electron-beam-induced defect chemistry, outlined in FIG. \ref{fig:Cartoon}. The as-synthesized NIMS-BN crystal has some initial concentration of vacancies and intercalated interstitials, illustrated in the cartoon in FIG. \ref{fig:Cartoon}(a). Electron beam induced diffusion increases the mobility of the interstitials causing the impurities to diffuse towards the naturally occurring vacancies within the material, resulting in the interstitial and vacancy combining into a defect complex\cite{dyck_placing_2017,hudak_directed_2018} (FIG. \ref{fig:Cartoon}(b)). We propose that the activation/deactivation behavior results from electron beam induced charge state switching of the color centers, similar to previous observations of NV centers in diamond\cite{gaebel_photochromism_2006}. In our model, the electron beam modifies the charge state of the defect causing it to change from a non-emissive to an emissive state, resulting in the production of photons (FIG. \ref{fig:Cartoon}(c)). The charge state of the defect can also be modified in the opposite sense, resulting in emitters switching on and off during a measurement. We do not see any evidence of emitters switching or blinking to a different emission band in our experiments. This effect would be fairly obvious in the SIs and would appear as a distinct component, complementary to the one shown in FIG. \ref{fig:HSOverview}(c) and \ref{fig:HSOverview}(d). The spatial loading would appear as a collection of point-like emitters accompanied by a set of truncated disks (truncated in the opposite sense compared those in FIG. \ref{fig:HSOverview}(c)).
\par
This change in charge state has two possible origins. One scenario is that incident beam and secondary electrons are captured by the defect complex, resulting in a negative charge state. Alternatively, the incident or secondary electrons ionize the defect, resulting in a positive charge state. Currently, neither possibility can be excluded suggesting future experiments and calculations. Previous studies have shown that a low energy electron beam can cause diffusion of impurities or vacancies, vacancy-impurity defect reconstruction, and charge state switching in other wide band gap semiconductors. While speculative, our proposed model is consistent with previous findings of electron beam induced luminescent diamond NV centers\cite{schwartz_effects_2012}.   
\par
We propose that this emission has not been previously observed in h-BN for several reasons. Past studies of color centers in h-BN have typically not used NIMS-BN which has a unique synthetic origin, but rather commercially obtained h-BN. We surmise that the Ba-B-N solvent precursor used in the NIMS-BN synthesis could produce barium impurities in these samples, which would not be present in commercially obtained material synthesized using different growth precursors. Under the assumptions of our model, the new emission would not be observed in the absence of this unique impurity. PL studies of color centers in h-BN that do use NIMS-BN have used a 532 nm laser excitation to probe the sample and have focused on light emission in a high wavelength range. This excitation energy is too low to probe a state that emits at 435 nm. It is likely that the charge recombination dynamics in this regime are dominated by non-radiative transitions. It is possible that the 435 nm emission is only present using CL because of the large number of electron-hole pairs created per incident electron, as well as the high intensity of the electron probe. A final, more speculative reason that this emission has not been observed to date is the possibility that the particular defect complex responsible for the emission needs the high energy density of the electron beam in order to drive interstitial-vacancy recombination.
\par
Our experimental results of the spectral and spatial character of this new emitter in h-BN show many features observed in quantum emitters, suggesting that this new color center is a potential source of single photons. The emission is highly localized, spectrally pure, exhibits a ZPL with phonon replicas, and can be modified with an electron beam, all features of SPEs in other solid state systems. The bunching behavior seen in the $g^{(2)}\left(\tau\right)$ measurement, while not proof of single photon emission, is not inconsistent with electron beam induced light emission from an ensemble of quantum emitters. While all observations are $\textit{consistent with}$ the behavior of a two-level-like system capable of single photon emission, further study is required to unambiguously confirm the quantum nature of this emitter. 
\section{Conclusion}
We have identified a new color center unique to high-quality hexagonal boron nitride using cathodoluminescence in the SEM. The emission is peaked at 435 nm and has spectral characteristics indicative of weak lattice coupling. The electron beam activates and deactivates emission from point defects in the crystal. We propose that this emission originates from a barium atom interstitial impurity forming a defect complex with a vacancy driven by the energy of the electron beam. The charge state of this defect is changed by the electron beam resulting in the emitters switching on and off. 

\begin{acknowledgments}
This research was supported primarily by the Director, Office of Science, Office of Basic Energy Sciences, Materials Sciences and Engineering Division, of the U.S. Department of Energy under Contract No. DE-AC02-05-CH11231 within the sp2-Bonded Materials Program (KC-2207) which provided for design of the experiment, and collection and analysis of the CL data. Additional support was provided by the Director, Office of Science, Office of Basic Energy Sciences, Materials Sciences and Engineering Division, of the U.S. Department of Energy under Contract No. DE-AC02-05-CH11231 within the van der Waals Heterostructures Program (KCWF16), which provided for BN sample preparation. 
This work was also supported by the National Science Foundation under Grant No. DMR-1807233 which provided for TEM structure and impurity characterization; and under Grant No. 1542741 which provided for development of optical instrumentation. Work at the Molecular Foundry was supported by the Office of Science, Office of Basic Energy Sciences, of the US Department of Energy under Contract No. DE-AC02-05CH11231. SMG acknowledges support from a Kavli Energy Nano Sciences Institute Fellowship and an NSF Graduate Fellowship.
\end{acknowledgments}

\bibliography{hBN.bib}

\end{document}